\documentclass[11pt,twoside]{article}

\usepackage{asp2014}

\aspSuppressVolSlug
\resetcounters

\begin{document}

\title{
To clean or not to clean? Influence of pixel removal on event reconstruction using deep learning in CTAO
}

\author{Tom~Fran\c cois,$^1$, Justine Talpaert$^1$ and Thomas Vuillaume$^1$}
\affil{$^1$LAPP, Univ. Savoie Mont-Blanc, CNRS, Annecy, France}

\paperauthor{Tom Fran\c cois}{tom.francois@lapp.in2p3.fr}{0000-0001-5226-3089}{LAPP, Univ. Savoie Mont-Blanc, CNRS}{}{Annecy}{}{}{France}
\paperauthor{Justine Tal}{justine.talpaert@lapp.in2p3.fr}{}{LAPP, Univ. Savoie Mont-Blanc, CNRS}{}{Annecy}{}{}{France}
\paperauthor{Thomas Vuillaume}{thomas.vuillaume@lapp.in2p3.fr}{0000-0002-5686-2078}{LAPP, Univ. Savoie Mont-Blanc, CNRS}{}{Annecy}{}{}{France}



\begin{abstract}
The Cherenkov Telescope Array Observatory (CTAO) is the next generation of ground-based observatories employing the imaging air Cherenkov technique for the study of very high energy gamma rays.
The software Gammalearn proposes to apply Deep Learning as a part of the CTAO data analysis to reconstruct event parameters directly from images captured by the telescopes with minimal pre-processing to maximize the information conserved. 
In CTAO, the data analysis will include a data volume reduction that will definitely remove pixels. This step is necessary for data transfer and storage but could also involve information loss that could be used by sensitive algorithms such as neural networks (NN).
In this work, we evaluate the performance of the $\gamma$-PhysNet~\citep{jacquemont2021deep} when applying different cleaning masks on images from Monte-Carlo simulations from the first Large-Sized Telescope.
This study is critical to assess the impact of pixel removal in the data processing, mainly motivated by data compression. 
\end{abstract}



\section{Introduction}
Gamma ray astronomy studies violent astrophysical phenomena (supernova remnants, gamma-ray bursts, active galactic nuclei, etc.) by detecting high-energy photons, called gamma rays.
When a gamma ray enters the atmosphere, it triggers a particle shower that can be detected by ground-based telescopes called Imaging Atmospheric Cherenkov Telescopes (IACTs).
The Cherenkov Telescope Array Observatory (CTAO), composed of multiple telescopes, provides an energy coverage from 20 GeV to more than 300 TeV.
Although the project is currently in the construction phase, the first Large-Sized Telescope (LST-1) prototype is operational under commissioning and has already made its first detections~\citep{lopez2021physics, abe2023multiwavelength}.
The data analysis can be performed using machine learning~\citep{fiasson2010optimization} or deep learning methods~\citep{de2023deep, jacquemont2021deep, parsons2020background, miener2021iact}. 
The software Gammalearn\footnote{\url{https://purl.org/gammalearn}}, based on $\gamma$-PhysNet~\citep{jacquemont2021deep} architecture that use Convolutional Neural Network (CNN) and attention mechanism, has been proposed to tackle the event reconstruction using observation images at Data Level 1 (DL1).
This method improves the reconstruction compared to traditional machine learning method but presents an extra sensitivity to changes in observation conditions~\citep{vuillaume2021analysischerenkovtelescopearray}.

Pixel removal is a classical step of IACTs data analysis, either to reduce the volume of data stored (DVR) or to remove the background noise in images before analysing the remaining signal. This step can have an important effect on CNNs by changing the image structure or by removing part of the signal. 

We present here the impact of pixel removal on the event reconstruction realised by the $\gamma$-PhysNet~ network. The study includes three cleaning methods compared with the analysis of raw images.

\articlefigure[width=0.7\textwidth]{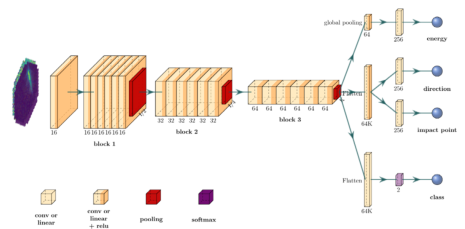}{fig:gammaphysnet}{$\gamma$-PhysNet~\citep{jacquemont2021deep} architecture: multi-task event reconstruction}

\section{Methods}

We distringuish 3 pixel selection methods:\\
\textbf{tailcut cleaning}: It consists in a two-threshold tail-cuts procedure as implemented in the \texttt{ctapipe} method~\citep{ctapipe-icrc-2023}. The method has 3 parameters : \textit{picture threshold}, which is the threshold above which all pixels are retained, \textit{boundary threshold}, which is the threshold above which pixels are retained if they have a neighbor already above the picture threshold. Finally, the last parameter is the minimum number of pixel neighbors, after the threshold steps, each pixel should have to survive the cleaning. The selected parameters for this work are $\{8,4,2\}$ respectively.\\
\textbf{lstchain cleaning}: This mask is used by default when analysing the DL1 Monte Carlo (MC) and real data from the LST-1. It is obtained with a \textit{tailcut} $\{8,4,2\}$ followed by a time consistency filtering, using the time map channel, as described in~\citep{abe2023observations,lstchain_adass_2020}.\\
\textbf{data volume reduction} (DVR): This method is similar to the pre-processing used in production on the LST-1 at La Palma. Here we apply a \textit{tailcut} with parameters $\{8,4,0\}$ followed by 3 dilations (morphological operation). The last parameter controls how many neighbour pixels are required to be kept. When this parameter equals 0, all isolated pixels are kept. The successive dilations allows to capture the neighbourhood of these pixels. 

We compare the three masking methods with the classical approach, named \textit{no mask}. The \textit{DVR} method is the more conservative method. If all or most of the signal is kept by this operation, we expect its reconstruction performances to match the \emph{no mask} ones.
Applying the different mask methods will set some pixel values to 0 in the images (see Fig. 1). On average, the DVR will remove 88\% of the pixels, and 98\% for \textit{tailcut} and \textit{lstchain} cleanings. If this operation removes all the pixels of the image, the associated event is discarded from the dataset. This is the case for 20\% of the images for \textit{lstchain cleaning} and \textit{tailcut cleaning}.

\articlefigure[width=0.7\textwidth]{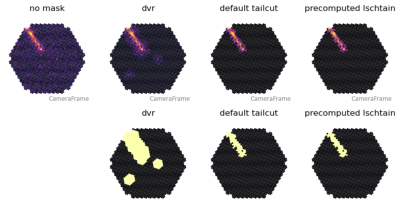}{fig:masks}{Visualisations of the binary masks application used in the cleaning study. Top row : input images after mask application. (Bottom-row) respective binary mask (yellow pixels are kept).}

\subsection*{Dataset}
The study has been conducted on Monte Carlo (MC) simulations Data Level 1 (DL1), tuned for the Crab Nebula source. This means the simulation has been conducted with extra noise to replicate the effect of a relatively high level of Night Sky Background (NSB). For each masking method, we use both protons and gammas with a 20 degrees zenith. For training, we use gamma-diffuse events, coming from all directions and, for inference, we use point source gammas. 

\section{Results}

We compare the three mask methods against the no mask one with Instrument Response Functions (IRFs), computed with glearn\_irfs\footnote{\url{https://gitlab.in2p3.fr/gammalearn/glearn_irfs}}. 
Event reconstruction is evaluated by 4 different metrics: energy resolution and bias to assess the energy regression, and angular resolution to assess the incident particle direction regression, and Area Under Curve (AUC) of the classification gamma against proton. 
Figure~\ref{fig:irfs} shows all of these metrics depending on the energy bins. 

\emph{DVR} and \emph{no mask} have a very similar response for every metric, showing that the DVR preprocessing is not affecting the $\gamma$-PhysNet behavior. 
On the other hand, the lstchain cleaning and tailcut cleaning present a significant drop in the performances. 
This is quite noticeable in the lower energy bins, below 100 GeV in particular.

\articlefigure[width=0.98\textwidth]{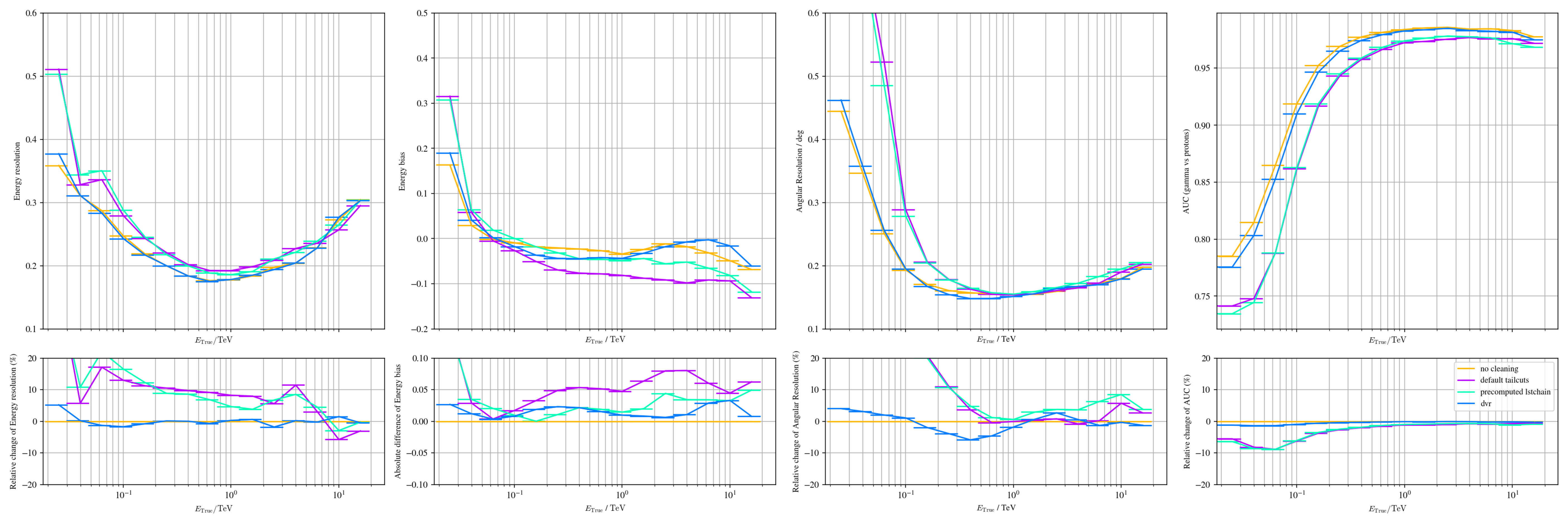}{fig:irfs}{
IRFs for the cleaning methods. From left to right: Energy resolution, Energy bias, Angular resolution and AUC. Top row represents the absolute plots and bottom row the relative plots against \emph{no mask}.
}

We also conducted the same experiments on MC generated with a lower amount of noise. We observe the same behavior.

\section{Conclusion}

The study shows that DVR preprocessing has very limited impact on $\gamma$-PhysNet~ performance demonstrating that the pre-processing used in production for LST-1 should have no impact on the analysis pipeline. Meanwhile, the other methods present a strong impact on the IRFs. In future work, we will study the impact of the cleaning on real data from the Crab Nebula.

\acknowledgements We gratefully acknowledge financial support from the agencies and organizations listed here: https://purl.org/gammalearn/acknowledgements and here: https://www.ctao.org/for-scientists/library/acknowledgments/

\bibliography{P210}  

\begin{thebibliography}{}
\expandafter\ifx\csname natexlab\endcsname\relax\def\natexlab#1{#1}\fi
\expandafter\ifx\csname url\endcsname\relax
  \def\url#1{\texttt{#1}}\fi
\expandafter\ifx\csname urlprefix\endcsname\relax\def\urlprefix{URL }\fi
\providecommand{\eprint}[2][]{\url{#2}}

\bibitem[{Abe et~al.(2023{\natexlab{a}})Abe, Abe, Abe, Aguasca-Cabot, Agudo,
  Crespo, Antonelli, Aramo, Arbet-Engels, Arcaro et~al.}]{abe2023observations}
Abe, H., Abe, K., Abe, S., Aguasca-Cabot, A., Agudo, I., Crespo, N.~A.,
  Antonelli, L., Aramo, C., Arbet-Engels, A., Arcaro, C., et~al.
  2023{\natexlab{a}}, The Astrophysical Journal, 956, 80

\bibitem[{Abe et~al.(2023{\natexlab{b}})Abe, Aguasca-Cabot, Agudo, Crespo,
  Antonelli, Aramo, Arbet-Engels, Artero, Asano, Aubert
  et~al.}]{abe2023multiwavelength}
Abe, S., Aguasca-Cabot, A., Agudo, I., Crespo, N.~A., Antonelli, L., Aramo, C.,
  Arbet-Engels, A., Artero, M., Asano, K., Aubert, P., et~al.
  2023{\natexlab{b}}, Astronomy \& astrophysics, 673, A75

\bibitem[{De et~al.(2023)De, Maitra, Rentala, \& Thalapillil}]{de2023deep}
De, S., Maitra, W., Rentala, V., \& Thalapillil, A.~M. 2023, Physical Review D,
  107, 083026

\bibitem[{Fiasson et~al.(2010)Fiasson, Dubois, Lamanna, Masbou, \&
  Rosier-Lees}]{fiasson2010optimization}
Fiasson, A., Dubois, F., Lamanna, G., Masbou, J., \& Rosier-Lees, S. 2010,
  Astroparticle Physics, 34, 25

\bibitem[{Jacquemont et~al.(2021)Jacquemont, Vuillaume, Benoit, Maurin, \&
  Lambert}]{jacquemont2021deep}
Jacquemont, M., Vuillaume, T., Benoit, A., Maurin, G., \& Lambert, P. 2021, in
  International Conference on Pattern Recognition (Springer), 174

\bibitem[{Linhoff et~al.(2023)Linhoff, Beiske, Biederbeck, Fröse, Kosack, \&
  Nickel}]{ctapipe-icrc-2023}
Linhoff, M., Beiske, L., Biederbeck, N., Fröse, S., Kosack, K., \& Nickel, L.
  2023, in Proceedings, 38th International Cosmic Ray Conference, vol. 444-703

\bibitem[{Lopez-Coto et~al.(2021)Lopez-Coto, Moralejo, Artero, Baquero,
  Bernardos, Contreras, Di~Pierro, Garc{\'\i}a, Kerszberg, L{\'o}pez-Moya
  et~al.}]{lopez2021physics}
Lopez-Coto, R., Moralejo, A., Artero, M., Baquero, A., Bernardos, M.,
  Contreras, J., Di~Pierro, F., Garc{\'\i}a, E., Kerszberg, D., L{\'o}pez-Moya,
  M., et~al. 2021, arXiv preprint arXiv:2109.03515

\bibitem[{L\'opez-Coto et~al.(2022)}]{lstchain_adass_2020}
L\'opez-Coto, R., et~al. (CTA, LST Project) 2022, in ASP Conf. Ser., vol. 532,
  357

\bibitem[{Miener et~al.(2021)Miener, L{\'o}pez-Coto, Contreras, Green, Green,
  Mariotti, Nieto, Romanato, \& Yadav}]{miener2021iact}
Miener, T., L{\'o}pez-Coto, R., Contreras, J., Green, J., Green, D., Mariotti,
  E., Nieto, D., Romanato, L., \& Yadav, S. 2021, arXiv preprint
  arXiv:2112.01828

\bibitem[{Parsons \& Ohm(2020)}]{parsons2020background}
Parsons, R.~D., \& Ohm, S. 2020, The European Physical Journal C, 80, 1

\bibitem[{Vuillaume et~al.(2021)Vuillaume, Jacquemont, de~Bony~de Lavergne,
  Sanchez, Poireau, Maurin, Benoit, Lambert, Lamanna, \&
  Project}]{vuillaume2021analysischerenkovtelescopearray}
Vuillaume, T., Jacquemont, M., de~Bony~de Lavergne, M., Sanchez, D.~A.,
  Poireau, V., Maurin, G., Benoit, A., Lambert, P., Lamanna, G., \& Project,
  C.-L. 2021, Analysis of the cherenkov telescope array first large-sized
  telescope real data using convolutional neural networks. \eprint{2108.04130},
  \urlprefix\url{https://arxiv.org/abs/2108.04130}

\end{thebibliography}


\end{document}